\documentclass [11pt]{article}         
\usepackage {apalike}

\begin{document}

\title{ Absolute Objects and  Counterexamples: Jones-Geroch Dust, Torretti Constant Curvature, Tetrad-Spinor, and Scalar Density} 

\author{J. Brian Pitts\footnote{ History and Philosophy of Science Graduate Program,
346 O'Shaughnessy, 
University of Notre Dame, 
Notre Dame, Indiana 46556 USA   
	email jpitts@nd.edu  	}   } 


\date{\today}
\maketitle


\abstract{ James L. Anderson analyzed the novelty of Einstein's theory of gravity as  its lack of ``absolute objects.'' Michael Friedman's related work has been criticized by Roger Jones and  Robert Geroch  for implausibly admitting as absolute the timelike 4-velocity field of dust in cosmological models in Einstein's theory.   Using the Rosen-Sorkin Lagrange multiplier trick, I complete Anna Maidens's argument that the  problem is not solved by prohibiting variation of absolute objects in an action principle.  Recalling Anderson's proscription of ``irrelevant'' variables, I generalize that proscription to locally irrelevant variables that do no work in some places in some models.  This move  vindicates Friedman's intuitions and removes the Jones-Geroch counterexample:  some regions of some models of gravity with dust are dust-free and so naturally lack a timelike 4-velocity, so diffeomorphic equivalence to (1,0,0,0) is spoiled.  Torretti's example involving constant curvature spaces is shown to have an absolute object on Anderson's analysis, \emph{viz.}, the conformal spatial metric density.  The previously neglected threat of an absolute object from an orthonormal tetrad used for coupling spinors to gravity appears resolvable by eliminating irrelevant fields.  However, given Anderson's definition, GTR itself has an absolute object (as Robert Geroch has observed recently):  a change of variables to a conformal metric density and a scalar density shows that the latter is absolute.
 }  

keywords:  absolute object, general covariance, spinor, tetrad, unimodular, density


\section{Introduction}

James L. Anderson analyzed the novelty of Einstein's so-called General Theory of Relativity (GTR) as its lacking ``absolute objects'' \cite{AndersonCoordinates,Anderson,AndersonGRG}.  Metaphorically, absolute objects are often described as a fixed stage on which the dynamical actors play their parts.  A  review  of Anderson's  definitions will be useful. Absolute objects are to be contrasted with dynamical objects. The values of the absolute objects do not depend on the values of the dynamical objects, but the values of the dynamical objects do depend on the values of the absolute objects \cite[p. 83]{Anderson}.  Both absolute objects and dynamical objects are, mathematically speaking, geometrical objects or parts thereof; the importance of this requirement will appear later.   It will be shown that Anderson's definition is naturally amended to avoid the Jones-Geroch dust counterexample more or less as Friedman envisioned.  The hitherto-unnoticed fact that Anderson's analysis detects an absolute object in Torretti's constant curvature spaces example blunts its force noticeably.  A previously unnoticed counterexample involving spinor fields is proposed and tentatively resolved using an alternative spinor formalism.  As a referee (who turns out to be Robert Geroch) suggested, however, GTR actually has an absolute object, using Friedman's definition and an attractive choice of variables.  Whether it is better to revise the notion of absolute object or revise that claim that GTR lacks them is presently unclear.

Before absolute objects can be defined, the notion of a covariance group must be outlined.  Here it will prove helpful to draw upon the unjustly neglected work of  Kip Thorne, Alan Lightman, and David Lee (TLL) \cite{TLL}; a useful companion paper (LLN) was written by   Lee, Lightman and W.-T. Ni \cite{LLN}. The TLL definition differs slightly from Anderson's in its notion of faithfulness.  According to TLL, 
\begin{quote}
A group $\mathcal{G}$ is a covariance group of a representation if (i) $\mathcal{G}$ maps [kinematically possible trajectories] of that representation into [kinematically possible trajectories]; (ii) the [kinematically possible trajectories] constitute ``the basis of a faithful representation of $\mathcal{G}$'' (i.e., no two elements of $\mathcal{G}$ produce identical mappings of the [kinematically possible trajectories]); (iii) $\mathcal{G}$ maps [dynamically possible trajectories] into [dynamically possible trajectories]. \cite[p. 3567]{TLL} \end{quote}

One can now define absolute objects.  They are, according to Anderson, objects with components $\phi_{\alpha}$ such that 
\begin{quote}
	(1) The $\phi_{\alpha}$ constitute the basis of a faithful realization of the covariance group of the theory.
	(2) Any $\phi_{\alpha}$ that satisfies the equations of motion of the theory appears, together with all its transforms under the covariance group, in every equivalence class of [dynamically possible trajectories]. \cite[p. 83]{Anderson}  
\end{quote}
Thus the components of the absolute objects are the same, up to equivalence under the covariance group,\footnote{There seems to be no compelling reason to require a covariance group instead of a mere covariance groupoid, a structure that would be a group if it were meaningful to multiply every pair of elements.  Einstein's equations on a background space-time, once one imposes a consistent notion of causality, have a covariance groupoid that is not a group \cite{NullCones1}. 
 }  in every model of the theory.  It is the dynamical objects that distinguish the different equivalence classes  of the dynamically possible trajectories \cite[p. 84]{Anderson}. One notices that the components of the absolute object need be the same, up to equivalence under the covariance group, for all \emph{dynamically} possible trajectories, not all \emph{kinematically} possible trajectories.  Might this matter have gone otherwise?  For most purposes this choice makes no difference, because typically those objects whose components are the same for all dynamically possible trajectories share the same feature for all kinematically possible trajectories.  This condition fails, however, in the context of Rosen's and Sorkin's deriving the flatness of a metric using a variational principle with a Lagrange multiplier, as will appear below.

 It has been asserted that the novel and nontrivial sense in which GTR is generally covariant is its lack of absolute objects \cite{Anderson} or ``prior geometry'' \cite[pp. 429-431]{MTW}. John Norton  discusses this claim with some sympathy \cite{NortonPCGC,Norton,NortonStumble}, though technical problems such as the Jones-Geroch dust and Torretti constant spatial curvature counterexamples are among his worries \cite{Norton,NortonStumble}.  Anderson and Ronald Gautreau encapsulate the definition of an absolute object as an object that ``affects the behavior of other objects but is not affected by these objects in turn.'' \cite[p. 1657]{AndersonGautreau}   Depending on how one construes ``affects,'' this summary might be serviceable, but only if used very cautiously.  On other occasions absolute objects are said to ``influence'' dynamical objects but not \emph{vice versa} \cite[p. 169]{AndersonGRG}.  Such terminology echoes Einstein and implies that absolute objects violate what Anderson calls a ``generalized principle of action and reaction'' \cite[p. 339]{Anderson} \cite[p. 169]{AndersonGRG}.  Norton has argued, rightly I think, that such a principle is hopelessly vague and arbitrary and that it should not be invoked to impart a spurious necessity to the  contingent truth that our best current physical theory lacks them \cite[pp. 848, 849]{Norton}.  One might also doubt whether terms such as ``affects,'' ``influence'' and ``act'' adequately capture what absolute objects typically do.  These terms suggest that the dynamical objects in question would have well-defined behavior if the absolute objects could somehow be `turned off,' so to speak (perhaps by replacing them with zero in the equations of motion), and that if the absolute objects were `turned on' again, they would alter the well-defined behavior of the dynamical objects in much the way that an applied electric field alters the motion of a charged particle.  But in important examples, such as Newtonian physics or special relativity, turning off many or all of the absolute objects destroys the theory: the equations of motion become degenerate or meaningless.  The absolute objects do not so much alter an otherwise happy situation as provide conditions in which the dynamical objects can have well-defined behavior.  Perhaps the stage metaphor for absolute objects is deeper than it seemed: presumably actors could put on a play on a stage consisting of a rubbery sheet or a giant pillow, or perhaps act in mid-air while falling freely,  but it is  easier to act on a firm wooden stage.  Thus the claim that absolute objects have some defect knowable \emph{a priori} easily may be taken too seriously. The fact that  it is even possible to do without them, as supposedly holds in Einstein's theory, should be something of a surprise (but instead turns out to be false, in light of the scalar density counterexample, on Friedman's definition).

In Anderson's framework, an important subgroup of a theory's covariance group is its symmetry group \cite[pp. 84-88]{Anderson}. One first defines the symmetry group of a \emph{geometrical object} as those transformations that leave the object unchanged.  If the transformations are  infinitesimal space-time mappings, then the Lie derivative of the geometrical object with respect to the relevant vector field vanishes for symmetries.  The symmetry group of a physical system or theory---Anderson makes no distinction between them here---is 
\begin{quote} 
	the largest subgroup of the covariance group of this theory, which is simultaneously the symmetry group of its absolute objects.  In particular, if the theory has no absolute objects, then the symmetry group of the physical system under consideration is just the covariance group of this theory. \cite[p. 87]{Anderson} \end{quote}  Thus, roughly speaking, the fewer absolute objects a theory has, the more of its covariance transformations are symmetry transformations.  For the example of a massive real scalar field obeying the Klein-Gordon equation in flat space-time in arbitrary coordinates, the covariance group is the group of diffeomorphisms, while the symmetry group is the 10-parameter Poincar\'{e} group corresponding to the ten Killing vector fields of Minkowski space-time.  For a massive real scalar field coupled to gravity in GTR, the covariance group is again the diffeomorphisms.  The symmetry group is also the diffeomorphisms, because any diffeomorphism leaves the set of absolute objects invariant, trivially, because there are no absolute objects (or so one thought until Geroch's scalar density counterexample appeared).  The fact that the space-time metric in GTR + massive real scalar field has no symmetries in general, though quite true, plays no explicit role in determining the symmetry group of the theory insofar as the space-time metric is dynamical rather than absolute.


Finding Anderson's definition obscure, Michael Friedman amended it in the interest of clarity \cite{FriedmanAbsolute,FriedmanFoundations}. Friedman takes his definition to express Anderson's intuitions, so the target of analysis is shared between them. As it turns out, Friedman has made a number of changes to Anderson's definitions, most of which seem to have received little comment by him or others,  so some comparison will be worthwhile.  

First, though Friedman's and Anderson's equivalence relations are laid out somewhat differently, a key difference between them is that Friedman's equivalence relation, which he calls $d$-equivalence, comprises only diffeomorphism  freedom \cite[pp. 58-60]{FriedmanFoundations}, not other kinds of gauge freedom such as local Lorentz freedom or electromagnetic or Yang-Mills gauge freedom, in defining the covariance group.  But local Lorentz freedom is a feature of the standard version of Einstein's GTR + spinors, for example. Anderson calls such groups besides diffeomorphisms ``internal groups'' \cite[pp. 35, 36]{Anderson}, though the term does not always fit perfectly for the examples available today.\footnote{In cases such as electromagnetic or Yang-Mills gauge freedom or local Lorentz invariance of an orthonormal tetrad, the name ``internal'' fits well, because the transformations happen independently at each space-time point. However, some symmetries that are not diffeomorphisms  resist being called internal.  One example is a theory with Einstein's equations formulated with a background metric \emph{tensor}.  Then there are two symmetries: diffeomorphisms and gauge transformations, both of which involve derivatives of the fields to arbitrarily high order \cite{Grishchuk,NullCones1} and so are nonlocal in their finite forms.  More famously, supersymmetry (which appears in supergravity and superstring theory) nontrivially combines internal and external symmetries.  Both examples became known after Anderson's work. The taxonomy of TLL \cite{TLL} is more capacious, but still does not comfortably accommodate Einstein's equations with a background metric.  As with covariance transformations, symmetry transformations can form a groupoid that is not a group \cite{NullCones1}. }  I find no argument for Friedman's restricting the relevant equivalence relation to diffeomorphisms, so perhaps he was unaware of this departure from Anderson's work. The goal is to distinguish physical sameness from conventional variation in descriptive fluff.  Because these other symmetries involve descriptive fluff as much as diffeomorphisms do, it seems that Anderson was more successful than Friedman on this point.  The role of internal groups in Anderson's work seems to have escaped Norton's notice \cite[pp. 847, 848]{Norton}.

Second, Friedman's mathematical language is less general than Anderson's and fails to accommodate some useful mathematical entities that Anderson's older component language permits.  Anderson, a working physicist, knows what sorts of mathematical structures physicists actually use and need, while Friedman restricts his attention to that narrower collection of entities that all modern coordinate-free treatments of gravitation or (pseudo-)Riemannian geometry presently  discuss, namely tensors and connections, but not, for example, tensor densities (especially of arbitrary real weight), which many such treatments neglect.  Considering how frequently other authors agree with Friedman's practice of neglecting tensor densities, it is worthwhile to recall how useful they are, if not essential in some applications.  In the literature on modern nonperturbative canonical quantization of gravity with Ashtekar's new variables and the like, tensor densities are used routinely.  Some authors write densities in a way that makes their weight manifest: a weight $2$ density has two tildes over it, a weight $-1$ density has a tilde below it, \emph{etc.}  Moreover,  the use of a densitized lapse function has proven useful in 3+1-dimensional treatments of the initial value problem\footnote{It is now customary in numerical general relativity to call the problem of inferring later or earlier states of a system from initial data the ``Cauchy problem,'' while the term ``initial value problem'' is reserved for the procedure of solving the constraint equations to get a set of initial data. This latter sort of problem exists only for constrained theories like GTR or Maxwell's electromagnetism.} in GTR  and the dynamical preservation of the constraint equations \cite{JantzenTaub,YorkHamilton}.  Perhaps these uses of densities are matters of convenience rather than necessary, because one can simulate tensor densities of integral weights using tensors.  However, this procedure is not so obviously available for most densities of non-integral weight;  it is generally unclear, for example, what a quantity with a third of an index or $\pi$ indices would mean.  But tensor densities of fractional weight have been  used in applications such as the conformal-traceless decomposition of Andr\'{e} Lichnerowicz and James York in solving GTR's initial value constraints in numerical general  relativity \cite{YorkCT,BrownCT}, unimodular variants of GTR (discussed in \cite{UnruhUGR,EarmanUGR}), and  quantum gravity \cite{PeresPolynomial,DeWitt67b,Leonovich}. 
Densities with \emph{irrational} weights are, if not essential, at least very useful in work on massive variants of Einstein's GTR \cite{OP,PittsMassive}. Thus Friedman's mathematical language does not accommodate these quantities that physicists use and perhaps require.  Hermann Weyl protested in 1920 against early clumsy efforts at component-free formalisms ``which are threatening the peace of even the technical scientist''  \cite[p. 54]{WeylSTM}.   Fortunately some modern authors have accommodated densities of arbitrary weight in a modern fashion\footnote{I thank referee Robert Geroch for emphasizing this point.  It is worth noting that Spivak's coordinate-free definition of arbitrary weight densities is spread out over pp. 314, 315, 391. One should also notice that tensor densities come in more than one kind; some can be of any real weight, while others are essentially of integral weight \cite{Golab,Spivak1}.  }
 \cite{Spivak1,Lang,Calderbank,Francaviglia2,FatibeneFrancaviglia}. 
Both the Torretti counterexample and the scalar density counterexample (discussed below) that finds an absolute object in GTR are most readily discussed  using  tensor densities.  Were tensor densities more widely discussed by philosophers of physics, likely Torretti's counterexample would not have been overestimated for so long, while the scalar density  counterexample would not have been overlooked for so long.  Anderson  did not neglect tensor densities, but simply erred in applying his definition of absolute objects to GTR by failing to consider the relevance of a simple  change of variables to irreducible geometric objects.  Thus we have examples of a problem noted  by M. Ferraris, M. Francaviglia and C. Reina: 
\begin{quote} In recent years, owing to their greater generality, geometric objects other than tensors began to enter physical applications, because in many cases using objects more general than tensors is essential [list of references omitted].  In fact, in spite of the widely known and systematic use of tensorial methods in mathematical physics, restricting ones [\emph{sic}] attention to tensors may often turn out to be misleading. \cite[p. 120]{FerrarisObject} \end{quote} Friedman's mathematical language is also inadequate to express the techniques used by V. I. Ogievetski\u{i} and I. V. Polubarinov in their atypical treatment of spinors coupled to gravity using a ``square root of the metric'' \cite{OPspinor}.  This spinor formalism should be useful in preventing the timelike leg of the orthonormal tetrad, which is typically used with spinors, from counting as an unwanted absolute object.

Third, while Friedman considers variously rich and spare versions of what is intuitively one theory (Newtonian gravity) and states a methodological preference for spare theories,  his treatment lacks the firm resolve of Anderson's demand that ``irrelevant'' variables be eliminated.  This requirement is also imposed  by TLL  \cite{TLL} and discussed by John Norton \cite{Norton}. One can readily adopt the Andersonian proscription of irrelevant variables to express Friedman's intuitions about ``natural'' choices of variables \cite[p. 59]{FriedmanFoundations} in relation to the Jones-Geroch dust counterexample.

A fourth  difference  pertains to the notion of standard formulations of a theory.  Anderson argues (somewhat confusingly) that theories should be coordinate-covariant under arbitrary manifold mappings; this move seems to be offered as a substantive claim rather than a conventional choice.  More understandably, TLL stipulate  that the standard form of a theory be manifestly coordinate-covariant.  Friedman, by contrast, takes as standard a form in which the absolute objects, if possible, have \emph{constant components} \cite[p. 60]{FriedmanFoundations}  and so have limited coordinate freedom.  Friedman implies  that  one can always choose coordinates such that the absolute objects (a) have constant components and (b) thus drop out of the theory's differential equations, which then pertain to the dynamical objects alone.  However, claim (a) is falsified by the counterexample of \mbox{(anti-)} de Sitter space-time as a background \cite{RosenBi78,LogunovConstant} for some \emph{specific} curvature value.  These space-times of constant curvature, at least for a fixed value of the curvature, satisfy Anderson's and Friedman's definitions of absolute objects for the space-time metric, but the components of the metric cannot be reduced to a set of constants.   An analogous example with spatial curvature is also available.    Anderson makes some effort to identify the `correct' or best formulation of a theory, a task taken up in more detail by TLL \cite{TLL}.  The latter authors' ``fully reduced generally covariant representation'' of a theory, unlike Friedman's ``standard formulation'' (p. 60), retains the full coordinate freedom by leaving the absolute objects as world tensors (or tensor densities, connections, or whatever  they are). Friedman's expectation that absolute objects be expressible using constant components is too strong to apply in every example.   Claim (b) is falsified by the example of massive versions of Einstein's theory 
\cite{OP,FMS,GrishchukMass,PittsMassive}. After a lull from the mid-1970s to the mid-1990s, massive variants of gravity have received considerable attention from physicists lately, especially particle physicists.  In those theories such that the background space-time metric is flat, its components  can be reduced to a set of constants globally by a choice of coordinates, but the background metric still does not disappear from the field equations because it appears in them algebraically, not merely differentially as Friedman apparently assumed tacitly.  Especially because (a) is false, the Thorne-Lee-Lightman fully reduced generally covariant formulation is therefore preferable to  Friedman's standard formulation, which  fails to exist in some interesting examples.  However, if one's goal is more historical, so that Newtonian gravity and special relativity without gravity are the main theories of interest, then Friedman's standard formulation suffices to illustrate the role of the Galilean and Poincar\'{e} groups, respectively.

Friedman's expectation that the components of absolute objects could be reduced to constants in general, though incorrect, usefully calls attention to the role (or lack thereof) of Killing vector fields and the like in analyzing absolute objects.    If the \mbox{(anti-)} de Sitter space-time examples show that constancy of components is too strict a criterion, the next best thing is to have a maximal set of 10 Killing vector fields in four space-time dimensions, whether commuting as in the flat space-time case or not as in the \mbox{(anti-)} de Sitter case.  One could generalize requirements on Killing vector fields in various ways \cite{ExactSolns}.   Because absolute objects need not be metric tensors, the general notion is not Killing vector fields, but generalized Killing vector fields, that is, fields such that the Lie derivative of the absolute objects vanishes.  Certainly some notion of constancy is one of the core intuitions that one has about absolute objects, though it plays no role in Anderson's definition of absolute objects, as John Earman has noticed \cite{Earman1974}. Newton's claim that absolute space ``remains similar and immovable'' is suggestive of symmetry within a model \cite{EarmanWorld}, not merely similarity between models.  Standard  examples of absolute objects usually have a fair number of generalized Killing vector fields.  In Anderson's terminology, most typical theories will have fairly large symmetry groups.  Usually at least a 7-parameter family of space- and time-translations and spatial rotations will be in the symmetry group, as in classical mechanics \cite{Goldstein}.  In GTR (including suitable matter fields), the lack (or scarcity, as the case may be) of absolute objects implies a  vast symmetry group.  This large group of all diffeomorphisms (or all volume-preserving ones) as symmetries of the absolute objects, in turn,  leads to an embarrassment of riches concerning local conservation laws, albeit noncovariant and not unique \cite[pp. 425, 426]{Anderson}.  From this fact follows the so-called nonlocalizability of gravitational energy.  If time translation invariance were required for absolute objects, then that criterion could exclude Norton's counterexample involving Robertson-Walker metrics \cite[p. 848]{Norton}.  The most typical and plausible examples of absolute objects do not apply forces that violate conservation laws; those that do, might well be called miraculous.  

\section{Confined objects and global space-time topology}

While absolute objects and dynamical objects are mutually exclusive, it is useful to have the third category of ``confined'' objects as well \cite{TLL}; these three categories are mutually exclusive and exhaustive, evidently.  Some entities that seemed intuitively absolute but do not satisfy Anderson's definition fit into the category of confined objects.   ``The confined variables are those which do \emph{not} constitute the basis of a faithful representation of the [manifold mapping group]'' \cite[p. 3568]{TLL}, which means (p. 3567) that there  exist two distinct elements of the manifold mapping group that produce identical mappings of the confined `variables'. The requirement that absolute objects form a faithful realization of the theory's covariance group is something that TLL carry over from Anderson \cite[p. 83]{Anderson}, though they have different definitions of faithfulness \cite[p. 3577]{TLL}. To avoid confusion with philosophical terminology (as a referee urged), let us call these new things ``confined objects'' rather than ``confined variables.'' TLL list universal constants as examples of confined objects. Indeed it is clear that structures that do not change at all under coordinate transformations are confined objects.  Some other examples of things unaffected by coordinate transformations that come to mind include the identity matrix, the Lorentz matrix $diag(-1,1,1,1),$ fixed Dirac $\gamma^{\mu}$ matrices, Lie group structure constants, and  Oswald Veblen's ``numerical tensors'' (which, in Veblen's usage, included  tensor densities).  The numerical tensors are the Kronecker $\delta^{\mu}_{\nu}$ symbol, which is trivially a world tensor, and the Levi-Civita totally antisymmetric $\epsilon$ symbol with values $1,$ $-1,$ and $0$; these values are the components of both a contravariant tensor density of weight 1 and a covariant tensor density of weight -1 \cite{VeblenQuadratic,Anderson,Spivak1}. 
It has been suggested by Harvey Brown that  the signature of the metric is importantly like an  absolute object   \cite{Brown,Maidens}. If the signature were an absolute object in the strict sense, then GTR would have an absolute object, contrary to Anderson's diagnosis of the novelty of GTR (though that diagnosis will be imperiled below on other grounds).  Anderson's and Friedman's works  have no category for expressing this immutable, externally prescribed  nature of the metric signature, because absolute objects are supposed to be geometric objects (tensor fields and the like).   The fact that the spacetime metric signature is unaffected by diffeomorphisms suggests that it counts as a confined object in the richer TLL taxonomy.    Restricting ourselves to space-time theories as usual, another issue worthy of consideration is the global topology of spacetime, which sometimes has  been neglected (but see \cite{Hiskes,FriedmanFoundations,Earman1974} \cite[pp. 298, 299]{StachelBZ}).  The global topology of spacetime is certainly untouched by diffeomorphisms, so it might be treated as a confined object. 

\section{Jones-Geroch counterexample and Friedman's reply}

With a clear grasp of absolute objects in hand, one can now consider the  Jones-Geroch counterexample that claims that the 4-velocity of cosmic dust counts, absurdly, as an absolute object by Friedman's or Anderson's standards.   Friedman concedes some force to this objection made by Robert Geroch and amplified by Roger Jones, here related by Friedman:
\begin{quote}
\ldots [A]s Robert Geroch has observed, since any two timelike, nowhere-vanishing vector fields defined on a relativistic space-time are $d$-equivalent, it follows that any such vector field counts as an absolute object according to [Friedman's criterion]; and this is surely counter-intuitive.  Fortunately, however, this problem does not arise in the context of any of the space-time theories I discuss.  It could arise in the general relativistic theory of ``dust'' if we formulate the theory in terms of a quintuple $ \langle M, D,g,\rho,U \rangle $, where $\rho$ is the density of the ``dust'' and $U$ is its velocity field.  $U$ is nonvanishing and thus would count as an absolute object by my definition.  But here it seems more natural to formulate the theory as a quadruple $ \langle M, D,g,\rho U \rangle $ where 
$\rho U$ is the momentum field of the ``dust.'' Since $\rho U$ does vanish in some models, it will not be absolute.  (Geroch's observation was conveyed to me by Roger Jones, who also suggested the example of the general relativistic theory of ``dust.''\ldots) \cite[p. 59]{FriedmanFoundations}  \end{quote}  Here  $D$ is the torsion-free covariant derivative compatible with $g$. Other sources, including what Roger Jones reported hearing from Robert Geroch, indicate  a qualification to \emph{local} diffeomorphic equivalence of nonvanishing timelike vector fields \cite[pp. 167, 168]{Jones} \cite{JonesGR} \cite[p. 84]{Trautman} \cite[p. 18]{Wald} \cite[pp. 198-200]{DodsonPoston}.  In any case nothing in my argument will depend on global \emph{versus} merely local equivalence between arbitrary neighborhoods. Jones also distinguishes the local diffeomorphic equivalence of nonvanishing timelike vector fields, which holds in general, from the (local) diffeomorphic equivalence of their covariant derivatives of various orders, which typically does not hold.

Below I will argue that Friedman's response is nearly satisfactory, though it has two weaknesses as he expressed it.   First, the statement ``$\rho U$ does vanish in some models'' ought to have said ``$\rho U$ does vanish in some neighborhoods in some models'' to  show that he is considering only  genuine models of GTR + dust (in which dust vanishes in some neighborhoods in some models),  rather than some  models with (omnipresent?) dust and some degenerate models which nominally have dust but actually have no dust anywhere.  The latter would seem to be a cheat.  As it stands, the reader is left to wonder whether such a cheat is doing important work for Friedman (though John Norton correctly read Friedman's proposal as ``relying \ldots on the possibility that $\rho$ vanishes somewhere'' \cite[p. 848]{Norton}).  Clearly some models with dust have neighborhoods lacking dust, and it is these models which will prevent the dust $4$-velocity from constituting an absolute object.  Second, Friedman's unfortunate notation $\rho U$ suggests that the mass current density (which I will call $J^{\mu}$) is logically posterior to $\rho$ and an everywhere nonvanishing timelike $U^{\mu}.$ If so, then one has not eliminated the absolute object after all. If a timelike nowhere vanishing $U^{\mu}$ exists in the theory, then it is absolute even if $\rho U^{\mu}$ vanishes somewhere and so is not absolute. Thus the significance of Friedman's use of $\rho U^{\mu}$ is left obscure. Instead one can take $J^{\mu}$ to be the fundamental variable, while the timelike $ U^{\mu}$ is a derived quantity defined wherever $\rho \neq 0.$ Alternatively, one can take $ U^{\mu}$ to be meaningful everywhere (and perhaps primitive), but vanishing where there is no dust.   If Friedman had said that $J^{\mu}$ or $U^{\mu}$ ``does vanish in some neighborhoods  in some models,'' then these two infelicities would have been avoided.  Perhaps these expository imperfections  led Roberto  Torretti to judge Friedman's reply \emph{ad hoc}  \cite{TorrettiFriedman} and John Norton to call it ``a rather contrived escape'' \cite[p. 848]{Norton}.  Once these problems are removed, the merit of Friedman's intuition shines brightly.

 Below I shall review more  discussion of this counterexample in the philosophical literature. Various neglected items from the physics literature will shed  light on long-standing philosophical debates about absolute objects.  Using the term ``variational'' for objects which are varied in an action principle \cite{GotayIsenberg}, one can safely follow Anderson in making ``absolute'' and ``dynamical'' mutually exclusive, while leaving open the connection between absoluteness and nonvariationality.  It will be shown that there exist  theories with variational absolute objects, at least if one does not exclude Rosen's variational principle as somehow illegal.  Such a theory can be obtained using Rosen's trick to fulfill Maidens's claim that the absolute special relativistic metric could be obtained variationally.  However these theories arguably violate Anderson's demand to eliminate irrelevant variables.   A natural extension of the proscription of irrelevant variables serves to eliminate the Jones-Geroch counterexample: the dust $4$-velocity $U^{\mu}$ does not count as an absolute object for GTR + dust because $U^{\mu}$ does not exist where there is no dust.

\section{Hiskes's redefinition of absoluteness, Maidens's worry, and Rosen's answer in advance}

Anne  Hiskes proposed amending the definition of absolute objects so that no field varied in a theory's action principle would be regarded as absolute \cite{Hiskes}.  Such a move makes use of what \emph{prima facie} seems to be a true generalization about absolute and dynamical objects.  This intuition was shared by the master.  Anderson wrote: \begin{quote} In addition to the differences between absolute and dynamical objects discussed in Section 4-3 there is another important difference that appears to be characteristic of these two types of objects.  The equations of motion for the dynamical objects can often be derived from a variational principle, especially if these objects are fields.  On the other hand, it appears to be the case, although we can give no proof of the assertion, that the equations of motion for the absolute objects do not have this property.\ldots In the following discussion we will assume that the equations of motion for the dynamical objects of a theory follow from a variational principle and that those for the absolute elements do not. \cite[pp. 88, 89]{Anderson}  \end{quote}  Thus Anderson suspected that  most or all dynamical objects are variational, while no absolute object is variational. Similar intuitions are manifest in the TLL and LLN papers \cite{TLL,LLN}.  Such a requirement also appears in their notion of being ``Lagrangian-based'' \cite[p. 3573]{TLL}. Recently John Earman has found it convenient to use ``absolute'' to \emph{mean} non-variational  \cite{EarmanUGR}.
 Anderson  was quite sensitive to the possibility of reformulating what intuitively seems like the same theory using various different sets, and indeed increasingly large sets, of variables in an action principle \cite[section 4.2]{Anderson}. Unlike Hiskes, he strove to define a unique correct formulation that gave the expected answers.    

 More recently, Anna Maidens has entertained the idea that Hiskes's redefinition could be deployed to remove the Jones-Geroch counterexample \cite{Maidens}.  If absolute objects must be nonvariational, while the dust $4$-velocity is variational, then the dust $4$-velocity is not absolute.  Following Hawking and Ellis \cite{HawkingEllis}, Maidens indicates how the equations for the timelike vector field can be derived from a variational principle.\footnote{One notices that Hawking and Ellis use a fluid variational principle with constrained variations, not the more familiar unconstrained variations.  In some respects this is a disadvantage, though Schutz and Sorkin observe that  it keeps one closer to the physical variables \cite{SchutzSorkin}.  They also observe that in many cases, including this one, one can eliminate the constraints on the variation (not to be confused with constraints in the sense of gauge theories \cite{Sundermeyer}) using Lagrange multipliers.  It seems to me that John Ray's variational principle might be preferable in the present context, because it involves varying $U^{\mu}$ itself and uses unconstrained variations \cite{RayFluid}.}  However Maidens is also sensitive to the large variety of choices of variables and even the number of field components in an action principle for what intuitively counts as a single theory.  Thus she expected such a use of Hiskes's redefinition to fail, because it eliminates the Jones-Geroch counterexample at the cost of introducing a new one.  More  specifically, Maidens has suggested that there might be some way to reformulate special relativistic theories such that the flat metric, which surely ought to count as absolute, is varied in the action principle.  If that could be done, then Hiskes's definition of absolute objects would prove to be too strict (the opposite problem from what the Jones-Geroch example suggests about Friedman's), because it fails to count the metric tensor of special relativity as an absolute object. (Maidens presumably should envision a weakly generally covariant formulation of special relativity,  though her notation is far from clear on that point.)  ``At this stage, however, we find a fly in the ointment, for its turns out that given suitable starting assumptions we can derive the Lorentz metric from an action principle.'' \cite[p. 262]{Maidens}  Supporting such claims would involve actually displaying a suitable Lagrangian density whose Euler-Lagrange equations give the desired results or else citing a source where such work had been done.   Surprisingly, she fails to do either one.  Success would involve finding an action principle for which the flatness of the metric holds for \emph{all} models (her case (c)), not just some (her case (a), p. 265).  A bit later she finds that ``it is an open question as to whether the metric of special relativity is derivable from an action principle.'' (p. 266)  Two pages later she once again claims that ``some of the physically necessary fixed background, e.g. the Lorentz metric, can also be derived from an action principle.'' (p. 268)  It is not easy to harmonize these fluctuating statements.  

Fortunately Maidens's expectation that the flatness of a metric (for all models) can be derived from a variational principle is in fact correct. The question was resolved by  Nathan Rosen in the 1960s \cite{RosenMultiplier,Rosen73}. He used an action principle involving a Lagrange multiplier field with 20 components, a trick recently used also by Rafael Sorkin  \cite{SorkinScalar}.  Thus  requiring absolute objects to be nonvariational gives an excessively strict definition, so the Jones-Geroch counterexample is not adequately addressed thereby.  Some objects that should count as absolute can be variational, as Maidens expected.  Rosen includes the following term in an action principle (after a change in notation to $\eta_{\mu\nu}$ for the metric in question, which is \emph{a priori} arbitrary apart from the signature) to force  $\eta_{\mu\nu}$ to be flat:
\begin{equation}
S = \int d^{4}x \sqrt{-\eta} R_{\rho\mu\nu\sigma}[\eta] P^{\rho\mu\nu\sigma}. \end{equation} This term is intended as a supplement to the action for a special relativistic theory, within which now  $\eta_{\mu\nu}$ would be subject to variation as well. $P^{\rho\mu\nu\sigma},$ a tensor with the same symmetries as the Riemann tensor for $\eta_{\mu\nu},$ serves as a Lagrange multiplier. Varying $P^{\rho\mu\nu\sigma}$ immediately yields the flatness of $\eta_{\mu\nu}.$ Varying $\eta_{\mu\nu}$ takes more work and gives an equation of motion especially involving the second derivatives of $P^{\rho\mu\nu\sigma}$.  That equation is not needed here.  Rosen seems to make secret use of the Euler-Lagrange equations from $P^{\rho\mu\nu\sigma}$  to discard terms involving $R_{\rho\mu\nu\sigma}[\eta]$ in his equations 10, 11, and 12; if so, then  his equation 12 is not an ``identity'' as he claims.  Alternately, he might be taking the metric to be flat before the variation but curved after it, as Sorkin proposes \cite{SorkinScalar}, if that is an intelligible alternative.\footnote{ It is perhaps worth noting that varying $P^{\rho\mu\nu\sigma}$ gives an equation of motion for $\eta_{\mu\nu}$ and varying $\eta_{\mu\nu}$ gives an equation of motion primarily for $P^{\rho\mu\nu\sigma}.$ Thus one should avoid expressions like ``the equations of motion for $\eta_{\mu\nu}$'' or ``the equations of motion for  $P^{\rho\mu\nu\sigma}$'' due to their ambiguity. }  As was noted above, Anderson's requiring component equality (up to equivalence under the covariance group) only for \emph{dynamically} possible trajectories is relevant here.  Using Rosen's trick, one has a geometric object such that its components agree for dynamically possible trajectories (``on-shell,'' as physicists say) but not for kinematically possible trajectories (``off-shell''), because the metric is not flat for all kinematically possible trajectories.

Anderson briefly states that one must remove irrelevant variables from the theory under analysis. He writes:
\begin{quote}
It is possible that a subset of the components of the [geometrical object characterizing the kinematically possible trajectories of the theory] do not appear in the equations of motion for the remaining components and furthermore can be eliminated from the theory without altering the structure of its equivalence classes.  Such a subset is obviously irrelevant to the theory.  We shall assume, therefore, that no subset of the components of [that geometrical object] is irrelevant in this sense.'' \cite[p. 83]{Anderson}  \end{quote}  Likewise TLL exclude the category of irrelevant variables \cite[p. 3569]{TLL}.  Anderson observes that \begin{quote} one can always construct a hierarchy of theories all of which have the same equivalence-class structure in the sense that the equivalence classes of these theories can be put into one-to-one correspondence with each other.  Two theories of such a hierarchy will differ both with regard to the mathematical quantities that describe their respective [kinematically possible trajectories] and their respective covariance groups.  However, the set of mathematical quantities that describe the [kinematically possible trajectories] of a given theory in such a hierarchy will contain, as subsets, those of each theory that precedes it in the hierarchy.  Likewise, its covariance group will contain, as a subgroup, the covariance group of each preceding theory.\ldots

The question then arises as to which theory of a hierarchy one should use to describe a given physical system.  The answer rests, of course, in the final analysis, on the measurements that one can make on the system.  It is necessary that each quantity used to describe the [kinematically possible trajectories] of a theory must, at least in principle, be measurable. \cite[p. 81]{Anderson}  \end{quote} Similar thoughts appear elsewhere in the text (pp. 306, 340).  This requirement of observability, an unfortunate whiff of verificationism, presupposes that all the physics resides in the field equations.\footnote{This last claim Anderson elsewhere implicitly appears to contradict when he considers boundary conditions (p. 75) and suggests (using ``furthermore'' on p. 83), surprisingly, that there could exist fields that do not appear in other fields' equations of motion, but which help to determine the structure of the theory's equivalence classes.  As it happens, recent work on field formulations of Einstein's equations provides an example: the flat metric does not appear essentially in the field equations, but it plays a role in the boundary conditions, topology, and the notion of gauge transformations \cite{NullCones1}. Boundary conditions are important in string theory as well \cite{StringBC}. Thus Anderson is overly hasty in eliminating the background metric after deriving Einstein's equations in flat space-time \cite[pp. 303-306]{Anderson}  in the fashion of Kraichnan \cite{Kraichnan}. While Kraichnan's use of a background metric in no way requires that quantization occur by covariant perturbation theory \cite{SolovevRTG}, historically the two projects have been linked in the minds of many.  Anderson critiqued perturbative approaches to Einstein's equations in response to a paper by Richard Arnowitt \cite{Arnowitt}.} But typically, fields that do useful work are observable, and Anderson's requirement of observability, if not entirely on target, at least emphasizes the importance of excluding idle  fields, such as $P^{\rho\mu\nu\sigma}$ appears to be.

While Rosen's trick vindicates Maidens's assertion that building nonvariationality into the notion of absolute objects is unsuccessful, Andersonian resources might be invoked  to exclude Rosen's trick as a form of cheating.  Anderson's prohibition  of irrelevant variables  appears to exclude theories making use of Rosen's trick,  because the dynamical evolution of the Lagrange multiplier $P^{\rho\mu\nu\sigma}$ has no effect on any other fields, whether gravitational or matter.   $P^{\rho\mu\nu\sigma}$ appears to do nothing useful  by Anderson's standards.  Making $\eta_{\mu\nu}$ variational and yet absolute could perhaps be useful in that it lets one treat the theory readily using the existing constrained dynamics formalism (\emph{e.g.}, \cite{Sundermeyer}), which has not made much room for nonvariational fields.  Making  $\eta_{\mu\nu}$  variational also allows one to define a conserved symmetric stress energy tensor without using the formal trick of the Rosenfeld approach, in which one replaces the flat metric by a curved one for taking a functional derivative and then restores flatness afterwards \cite{Deser}. 
  Whether Rosen's trick or Rosenfeld's is preferable is open to discussion, but an Andersonian elimination of the Lagrange multiplier field as irrelevant would be at least a defensible view.

Where does this dialectic  leave us?  Maidens proposed and rejected using Hiskes's redefinition of absolute objects to exclude the Jones-Geroch counterexample to Friedman's account of absolute objects.  Maidens's missing proof was supplied in advance by Rosen.  But Rosen's trick seems not to count against Anderson's version of the intuition that absolute objects are nonvariational, because Anderson wisely has criteria for eliminating irrelevant variables.  Does it follow that Anderson's intuition, in the larger context of his project that excludes irrelevant variables, is vindicated?  That is, if we accept Anderson's definitions and proscriptions, should we also accept his intuition that fields are variational if and \emph{only if} they are dynamical?  As it turns out, Anderson's generalization survives this alleged counterexample but might be threatened by another in which all fields are variational but there is still an absolute object.  I have in mind parametrized theories 
\cite{Sundermeyer,Kuchar73,Schmelzer,Arkani,NortonGC,EarmanUGR}, in which preferred coordinates are rendered variational.  One often calls the results ``clock fields.'' Perhaps some uses of clock fields could be excluded as irrelevant---not because the fields themselves are irrelevant, but because perhaps their variationality  is.  On the other hand, if clock fields are used to satisfy an appropriate notion of causality in bimetric theories like massive variants of Einstein's equations \cite{PittsMassive,NullCones1,Schmelzer}, then their variationality is relevant. Parametrized theories require more discussion than is appropriate here, however.  The scalar density example below is, at present, another example of a variational yet absolute object.  

\section{Eliminating local irrelevance excludes  the Geroch-Jones vector field}

If  Maidens's proposed and rejected use of  Hiskes's redefinition is set aside for violation of Anderson's prohibition of irrelevant variables, then the Jones-Geroch counterexample still remains to be addressed.  Now it turns out that Anderson's and TLL's proscription of irrelevant variables, if it does not quite remove the Jones-Geroch counterexample, at least inspires a  gentle amendment that does the job. This amendment seems especially appropriate after one notices that TLL replace \cite[p. 3566]{TLL}  Anderson's notion of geometrical object \cite[pp. 14-16]{Anderson} with Andrzej Trautman's notion of a geometric object \cite{Trautman}.  Presumably both notions aim to capture the same intuition.

Given the relative inaccessibility of Trautman's lectures, it will be worthwhile to quote his definition of geometric objects in detail:
\begin{quote} Let $X$ be an $n$-dimensional differentiable manifold.\ldots [S]ince tensors are not sufficient for all purposes in geometry and physics, [\emph{sic}] for example scalar densities are not tensors, to avoid having to expand definitions and theorems whenever we need a new type of entity, it is convenient to define a more general entity, the \underline{geometric object}, which includes nearly all the entities needed in geometry and physics, so that definitions and theorems can be given in terms of geometric objects so as to hold for all the more specialized cases that we may require.

Let $p \in X$ be an arbitrary point of $X$ and let $\{x^a\}, \{x^{a^\prime} \}$ be two systems of local coordinates around $p.$ A \underline{geometric object field} $y$ is a correspondence $$ y: (p, \{x^a\}) \rightarrow (y_{1}, y_{2},\cdot \cdot \cdot \; y_{N}) \in R^N$$ which associates with every point $p \in X$ and every system of local coordinates $\{x^a\}$ around $p$, a set of $N$ real numbers, together with a rule which determines $(y_{1^\prime}, \cdot \cdot \cdot \; y_{N^\prime})$, given by $$ y: (p, \{x^{a^\prime} \}) \rightarrow (y_{1^\prime}, \cdot \cdot \cdot \; y_{N^\prime}) \in R^N$$
 in terms of the  $(y_{1}, y_{2},\cdot \cdot \cdot \; y_{N})$ and the values of [\emph{sic}] $p$ of the functions and their partial derivatives which relate the coordinate systems $\{x^a\}$ and $ \{x^{a^\prime} \}.$\ldots  The $N$ numbers  $(y_{1}, \cdot \cdot \cdot y_{N})$ are called the \underline{components} of $y$ at $p$ with respect to the coordinates $\{x^a\}$. \cite[pp. 84, 85]{Trautman} \end{quote}  Trautman then notes that spinors are not geometric objects.  He also notes that some objects that are not themselves geometric objects are  nonetheless \emph{parts} of geometric objects.  \emph{Pace} Friedman's nonstandard usage \cite[p. 359]{FriedmanFoundations}, the class of geometric objects is not exhausted by tensors and connections. Trautman's definition was fairly typical in its time, though a bit streamlined for physicists' use.    Geometric objects were considered with great thoroughness by Albert Nijenhuis \cite{Nijenhuis}. A more recent treatment of them using modern differential geometry has been given by Ferraris, Francaviglia, and  Reina \cite{FerrarisObject}.


The reader will notice that Trautman's geometric objects are defined at every point in the space-time manifold.  That fact is of special relevance for the dust example, because it implies that if a dust $4$-velocity timelike unit vector field $U^{\mu}$ is used as a variable in the theory, then a dust $4$-velocity timelike unit vector must be defined at every point in every model, \emph{even if no dust exists in some neighborhoods in some models}.  Here one recalls Anderson's and TLL's call for the elimination of irrelevant variables; Friedman also recognizes the value of eliminating surplus structure.  It is not clear that existing notions of irrelevance apply strictly  to the present case. The dust $4$-velocity is locally irrelevant, not globally irrelevant, one might say. Perhaps  the authors had in mind fields that satisfy equations somewhat like the Klein-Gordon equation as their primary examples, as theoretical physicists often do; for such fields irrelevance is likely to be global.  But now that the question is raised, it does seem clear that wherever there is no dust, there ought not to be a dust $4$-velocity timelike unit vector either---at least not if the task at hand is testing theories for absolute objects.

There seem to be three initially plausible alternatives concerning the dust $4$-velocity where the dust has holes in some model.  First, one might retain a timelike $4$-velocity vector even in holes in the dust, while expecting the $4$-velocity values in the dust holes to be mere gauge fluff. It is noteworthy that at least some perfect fluid variational principles in the physics literature yield timelike unit vector $4$-velocities even where there is no fluid \cite{RayFluid}. Perhaps mathematical convenience commends this option, though I find that Ray's variational principle can be modified to lack a timelike $4$-velocity in holes in the fluid. 
 Presumably one could show that the value of a timelike $4$-velocity vector is in fact gauge fluff in dust holes by using the Dirac-Bergmann constrained dynamics technology \cite{Sundermeyer}, though one might run into technical challenges with changes of rank or with the noncanonical Poisson brackets that can appear in fluid mechanics \cite{Morrison}.  In any case, the timelike dust $4$-velocity in dust holes has no physical meaning, yet leads one to conclude that the theory has an absolute object.  Clearly any absolute object whose existence is inferred only by using physically meaningless quantities is spurious.  If one allowed physically meaningless entities into a theory while testing for absolute objects, then one could take any theory and construct an empirically equivalent theory with as many absolute objects as one wants. One could concoct a version of GTR with Newton's absolute space, for example. To permit such a procedure is just to give up Anderson's program of analyzing the uniqueness of GTR, because analysis involves \emph{trying} to get the intuitively known right answer as a consequence of some criteria.  Anderson and TLL call for the elimination of irrelevant variables in order to address  just this sort of problem.  One might call the entities that they reject ``globally irrelevant variables'' because such entities play no role at any space-time point in any model.  The Jones-Geroch example shows, I conclude, that one must also exclude  ``locally irrelevant variables,'' entities that play no role in some neighborhoods in some models.  One could consider whether mathematical entities that play no role at some space-time points or sets of measure zero should also be excluded as locally irrelevant, but there might be technical reasons for admitting them.  

The two remaining options avoid this spurious absolute object in different ways. One option is to take the mass current density $J^{\mu}$ to be the primitive variable and regard  $U^{\mu}$ and the dust density $\rho$ as derived.  Then  $\rho$ is defined  by $ \rho = \sqrt{ -J^{\mu}g_{\mu\nu}J^{\nu} }.$ The $4$-velocity $U^{\mu}$ is naturally defined by $$ U^{\mu} = \frac{ J^{\mu} }{ \sqrt{ -J^{\nu} g_{\nu\alpha} J^{\alpha} } } ,$$ so $U^{\mu}$ is only meaningful where the denominator $\rho$ is nonzero. That consequence is plausible on physical grounds and blocks the Jones-Geroch counterexample.  The theory is thus formulated using a quadruple $ \langle M, D,g, J \rangle,$ not Friedman's quadruple $ \langle M, D,g,\rho U \rangle $  or the  quintuple $ \langle M, D,g,\rho,U \rangle.$   In some models $J^{\mu}$  vanishes at some space-time points in some models of GTR + dust, so  $U^{\mu}$ is undefined in such cases.  Neither $J^{\mu}$ nor $U^{\mu}$ is a Gerochian nowhere vanishing timelike vector field for all models.  By contrast, the mass current density $J^{\mu}$, which is equal to $ \rho U^{\mu}$   where $\rho \neq 0,$ automatically vanishes where there is no dust and is continuous at the transition from dust to vacuum.  Thus Friedman's suggestion that it is more ``natural'' to use the mass current density, once freed from the two infelicities noted at the beginning, is seen to be very reasonable.   

 The other option is to take $U^{\mu}$ to be meaningful but vanishing in those places in certain models where the dust has holes.\footnote{ One need not commit oneself to $J^{\mu}$ as primitive and $U^{\mu}$ as derived. I am indebted to Don Howard for insightful probing about choices of primitive variables. If $U^{\mu}$ is allowed to vanish in some places, then it is not rightly everywhere called the dust $4$-velocity, as a  referee notes.}  Although $U^{\mu}$ exists everywhere, it vanishes in some places in some models, so not every neighborhood in every model has $U^{\mu}$ that is gauge-equivalent to $(1,0,0,0).$   Anderson's definition of absolute object requires that, for any component $\phi_{\alpha}$ of an absolute  object in a theory, ``[a]ny $\phi_{\alpha}$ that satisfies the equations of motion of the theory appears, together with all its transforms under the covariance group, in every equivalence class of [dynamically possible trajectories].'' \cite[p. 83]{Anderson}  Even if we drop Anderson's requirement of global equivalence in favor of Hiskes's (and Friedman's \cite[pp. 58-60]{FriedmanFoundations})
local equivalence, $U^{\mu}$ does not count as absolute.  In dust-filled regions in a model, the dust $4$-velocity $U^{\mu}$ is diffeomorphic (at least in a neighborhood) to $(1,0,0,0),$ but in dust holes $U^{\mu}$ is diffeomorphic to $(0,0,0,0)$ instead. Thus $U^{\mu}$, like $J^{\mu}$, is not an absolute object. One might tolerate as harmless the surplus structure embodied in the vanishing $U^{\mu}$ vectors, though the mathematical discontinuity of the vector field makes it difficult to defend this option on grounds of mathematical convenience.

 If one chooses to restrict one's attention to models of GTR + dust that do have dust everywhere and always, such gerrymandering is simply changing the subject to consider a different theory.  If one takes a semantic view of theories, then restricting attention to such a set of models is just to discuss some new theory besides GTR + dust, namely GTR + omnipresent dust. Manifestly GTR + omnipresent dust is a proper subset of GTR + dust.  GTR + omnipresent dust has the peculiar feature of  describing ``dust'' with such attributes as necessary existence, omnipresence and eternality, attributes more suited to a Deity than to dust.  Moreover, GTR + omnipresent dust is not the set of cosmological models of GTR.  For example, one can write down cosmological models in which dust is present but not omnipresent \cite[p. 166]{Feynman} \cite{KleinIsland,SmollerTemple}.  More realistic models  include eras of radiation domination and perhaps dark energy, so dust is not even a good description of matter in every region of space-time in cosmological models in GTR.  In short, GTR + omnipresent dust has no essential physical relevance to cosmology.  Having suitably deflated  expectations regarding the theory's physical import, one can proceed to test it for absolute objects.   The new theory GTR + omnipresent dust  has an absolute object.  But why shouldn't it?  Surely no one has well founded intuitions to the contrary. Any matter with the attributes of necessary existence, omnipresence and eternality just isn't  much like dust, but  rather has the vaguely theological flavor that both friends and foes of absolute objects (such as Newton and Einstein in his Machian aspect, respectively--if the reader will pardon the anachronism) have sensed.  Anderson anticipated the fact that one could consider a proper subset of models for which some field would count as absolute without counting as absolute for the full set of models.  He wrote:
\begin{quote}  We should perhaps emphasize that we are discussing here universal absolute objects, which must appear in the description of every [dynamically possible trajectory] of our space-time description.  It is quite possible that, for a subclass of [dynamically possible trajectories], one or more dynamical objects satisfy the criteria of Section 4-3 and so play the role of absolute objects for those [dynamically possible trajectories].\ldots The existence of such special subclasses of [dynamically possible trajectories] as those discussed above does not, of course, constitute a violation of the principle of general invariance as we have formulated it.  Only the existence of universal absolute objects would do so.  \cite[pp. 339, 340]{Anderson}  \end{quote}  Thus Anderson reminds us that absolute objects are universal, not (so to speak) provincial like the dust $4$-velocity.   While the dust $4$-velocity constitutes an absolute object for the theory GTR + omnipresent dust, it does not constitute an absolute object for GTR + dust due to the failure of universality.   Thus Friedman's intuition, as modified above, is vindicated.  The alleged Jones-Geroch counterexample fails to count as an absolute object for GTR + dust and thus fails to undermine Friedman's analysis after a slight amendment using Andersonian resources.

One might summarize Friedman's reply, as amended above, as follows:  Geroch's merely mathematical vector field is irrelevant and eliminable because it does no physical work, while Jones's dust application of the vector field  does physical work but violates the condition of being meaningful and everywhere nonvanishing in all models and so violates the diffeomorphic equivalence needed for absoluteness. At this stage a summary might be useful.  Physics literature previously unappreciated by philosophers of physics has been shown to shed light on the Jones-Geroch counterexample to Friedman's (and likely Anderson's or TLL's) definition of absolute objects. An old result from Rosen vindicates Maidens's claim that Hiskes's redefinition of absolute objects could not be used to eliminate the Jones-Geroch counterexample without generating a new counterexample.  The neglected  but valuable  paper by TLL and some infrequently attended  parts of Anderson's book proscribe irrelevant variables, a fact with important consequences.  This proscription perhaps can be used to exclude Rosen's trick for deriving flat space-time from a variational principle.  Then Anderson's generalization that absolute objects are variational and \emph{vice versa} would seem to be rehabilitated, at least provisionally, though the clock fields of parametrized theories pose further questions(as does the scalar density example below).  If variationality cannot be invoked to remove the Jones-Geroch counterexample, then some new move is required.  Again the Anderson-TLL proscription of irrelevant variables is helpful, in spirit if not in letter.  Excluding locally irrelevant values of the field $U^{\mu}$, which purports to be the $4$-velocity field of dust, would imply that $U^{\mu}$ is undefined wherever the dust vanishes, while the mass current $J^{\mu}$ vanishes there. Alternatively, $U^{\mu}$ and  $J^{\mu}$ both vanish there.  Either way, GTR + dust fails to have an everywhere nonvanishing timelike vector field that exists in all models.  Thus a slight amendment of the Anderson-Friedman tradition using the Andersonian opposition to irrelevant variables eliminates the Jones-Geroch counterexample.  

\section{Torretti's example of constant curvature spaces has  Andersonian absolute object}

A second long-standing worry concerning  the Anderson-Friedman absolute objects project was suggested by Roberto Torretti  \cite{TorrettiFriedman}.  He considered a theory of modified Newtonian kinematics in which each model's space has constant curvature, but different models have different values of that curvature.   Because every model's space has constant curvature, such a theory surely has something rather like an absolute object in it, Torretti's intuition suggests.  Though contrived, this example is relevantly like the cases of  de Sitter or anti-de Sitter background metrics of constant curvature that are sometimes discussed in the physics literature  (\emph{e.g.}, \cite{RosenBi78,LogunovConstant}), where one often lumps together space-times with different values of constant curvature. The failure of the  metrics to be locally diffeomorphically equivalent for distinct curvature values entails that the metric tensor does not satisfy Anderson's or Friedman's definition of an absolute object (or TLL's, for that matter).  Thus Torretti concludes that Anderson's project is not adequate for achieving the goals that Friedman has or ought to have.

How seriously one takes Torretti's objection will depend in part upon the degree that one shares Torretti's  expectations
for absolute objects.  Though Anderson evidently invented the term and defined it, Torretti expects a much broader array of applications that does Anderson.   The justice of this expectation depends on what sorts of claims Friedman made on behalf of the Anderson-Friedman project, as well as how seriously one takes Anderson's non-technical glosses about acting without being acted upon and the like.  A homely example will help. A lawn mower is a modest but nontrivial tool for caring for the grass in one's yard.  One can imagine a more impressive machine that also trims around obstacles and pulls weeds, though no such machine exists presently.  Anderson's project, like a lawn mower, is a tool that largely succeeds in satisfying a  modest but nontrivial goal.  Torretti is more ambitious in his goal, but his tool, like a lawn mower that also trims around edges and pulls weeds, does not presently exist.  In the absence of the more impressive tool, one might be content with the more modest tool that is  presently available.  It also seems peculiar that in Torretti's example,  the value of the curvature of space is  contingent (varying across models), but necessarily (in every model) the value at one moment is the same as that at another moment. Perhaps the failure of  an Anderson-Friedman definition  of absolute objects to count the metric as absolute in Torretti's example shows a quirk in the example rather than the definition.

Though neither Torretti nor later writers seem to have noticed, Anderson's analysis, when applied to Torretti's example, does yield a very specific and reasonable conclusion involving an absolute object.  Though the spatial metric is not absolute, the conformal spatial metric density, a symmetric $(0,2)$ tensor density of weight $-\frac{2}{3}$ (or its $(2,0)$ weight $\frac{2}{3}$ inverse) is an absolute object.  This entity, when its components are expressed as a matrix,  has unit determinant.  It appears routinely in the conformal-traceless decomposition used in finding initial data in numerical studies of GTR.  It defines angles and relative lengths of vectors at a point, but permits no comparison of lengths of vectors at different points. In three dimensions, conformal flatness of a metric is expressed by the vanishing of the Cotton tensor \cite{Aldersley,Macias}, not the Weyl tensor, which vanishes identically.  That the  conformal metric density is an absolute object is shown in the following way.  Every space with constant curvature is conformally flat  \cite{Wolf,RobertsonNoonan,MTW}.  For conformally flat spatial metrics, manifestly the conformal parts are equal in a neighborhood up to diffeomorphisms.  The conformal part just is the conformal metric density, so the conformal metric density is the same (within a diffeomorphism) locally for every model in Torretti's theory.  One could have the intuition that Anderson's analysis captures as  absolute  everything that it ought to capture.  I conclude that the force of this counterexample has been  overestimated.  Concerning  Norton's modification of Torretti's example to Robertson-Walker metrics \cite[p. 848]{Norton}, analogous comments could be made: these space-\emph{times} are conformally flat 
\cite{InfeldSchild,Tauber,Padmanabhan,Kuchowicz} and so have as an absolute object the space-time conformal metric density.  

\section{Tetrad-spinor: Avoiding absolute object by eliminating irrelevant fields }

One potential counterexample to the Anderson-Friedman example that seems not to have been noticed arises from the use of an orthonormal tetrad formalism, in which the metric tensor (or its inverse) is built out of four orthonormal vector fields $e^{\mu}_{A}$ by the formula $g^{\mu\nu}=e^{\mu}_{A} \eta^{AB} e^{\nu}_{B}$ or the like. Four vector fields have among them 16 components, rather more than the 10 components of the metric, so there is some redundancy that leaves a new local Lorentz gauge freedom to make arbitrary position-dependent boosts and rotations of the tetrad. It is unnecessary to use  a tetrad instead of a metric as the fundamental field when gravity (as described by GTR) is coupled to bosonic matter (represented by tensors, tensor densities or perhaps connections).  However, it is widely believed to be necessary to use an orthonormal tetrad to couple gravity to the spinor fields that represent electrons, protons, and the like  \cite{Weinberg,DeserVierbein,FatibeneFrancaviglia}. The threat of a counterintuitive  absolute object then arises.  Given both local Lorentz and coordinate freedom, one can certainly bring the timelike leg into the component form  $(1,0,0,0)$ at least in a neighborhood about any point. (Aligning the tetrad with the simultaneity hypersurfaces is known as imposing the time gauge on the tetrad 
\cite{DeserVierbein}.)  Unlike the dust case, there cannot be any spacetime region in any model such that the timelike leg of the tetrad vanishes.  Thus GTR coupled to a spinor field using an orthonormal tetrad gives an example of a Gerochian vector field: nowhere vanishing, everywhere timelike, gauge-equivalent  to $(1,0,0,0)$, and (allegedly) required to couple the spinor and gravity and thus not irrelevant. Like clock fields, the timelike tetrad leg also appears to be both variational and absolute.  If it is true that coupling spinors to gravity requires an orthonormal tetrad and that an orthonormal formalism for GTR yields an absolute object, then the intuitively absurd conclusion that GTR + spinors has an absolute object follows.

  Before discussing the tetrad-spinor issue, it is worthwhile to consider Anderson's treatment of spinors of the Dirac equation in a gravitational field (pp. 358-360).  Anderson entertains  the worry that $\gamma^{\mu}$ might be an absolute object in flat spacetime, in fact one with a symmetry group smaller than the Poincar\'{e} group (though in this context $\gamma^{\mu}$ is not a vector under \emph{arbitrary} coordinate transformations, so it is not eligible to be an absolute object by Anderson's standards, it would seem).  Turning to curved spacetime,  Anderson avoids using an orthonormal tetrad by using variable Dirac matrices $\gamma^{\mu}$ satisfying $\gamma^{\mu} \gamma^{\nu} + \gamma^{\nu} \gamma^{\mu} = 2g^{\mu\nu} I.$  What follows is a formalism with an internal symmetry group (apparently global) unrelated to the group of spacetime mappings.  However, the implicit relationship between $\gamma^{\mu}$ and $g^{\mu\nu}$ leaves obscure what a suitable action principle might be for deriving the Einstein-Dirac equations and what variables it would involve.  Thus one can hardly even test Anderson's formalism for absolute objects; his treatment of spinors is just incomplete.   By contrast the tetrad-spinor formalism avoids such difficulties.

  The tetrad-spinor example seems rather more serious a problem for definitions of absolute objects than the  Jones-Geroch cosmological dust example was, because the spinor field is surely closer to being a fundamental field than is dust or any other perfect fluid. Spinors (actually vector-spinors for spin $\frac{3}{2}$) are also required in supergravity, where internal and external symmetries are combined, not to mention (super)string theory.  On another occasion I expect to explain in more detail how to remove irrelevant variables here and thus avoid  this unexpected absolute object.  This removal is achieved using the alternative spinor formalism of V. I. Ogievetski\u{i}  and I. V. Polubarinov \cite{OPspinor}  to  eliminate ``enough''  of the orthonormal tetrad as irrelevant that the timelike nowhere vanishing vector field disappears from the theory. A brief summary suffices here.  Their formalism's ``square root of the metric'' resembles an orthonormal tetrad gauge-fixed to form a symmetric matrix by sacrificing the local Lorentz freedom while preserving diffeomorphism freedom.  The square root of the metric has only ten components rather than sixteen and can be computed using a binomial series expansion.

\section{Scalar density example and unimodular GTR: Does GTR lack absolute objects?  }

Unimodular GTR was invented by Einstein, was discussed by Anderson along with David Finkelstein \cite{AndersonFinkelstein}, and is rather well known today \cite{EarmanUGR}.  Still  it turns out that consideration of unimodular GTR helps one to reach the startling conclusion that not only it, but GTR itself, has an absolute object on Friedman's definition.  (While serving as a referee, Robert Geroch  proposed this counterexample, though using different mathematical variables.)   Unimodular GTR comes in two flavors: the coordinate-restricted version in which only coordinates that fix the determinant of the metric components matrix to $-1,$ and the weakly generally covariant version that admits any coordinates with the help of a nonvariational scalar density (usually of weight 1 or 2, but any nonzero weight suffices) and a dynamical conformal metric density, which is  a $(0,2)$ tensor density of weight $-\frac{2}{n}$ or a $(2,0)$ tensor density of weight $\frac{2}{n}$ in $n$ space-time dimensions.  As Anderson and Finkelstein observe, a metric tensor as a geometric object is reducible into a conformal metric density and a scalar density. They have in mind an equation along these lines:
\begin{equation}
g_{\mu\nu} = \hat{g}_{\mu\nu} \sqrt{-g} \, ^{ \frac{2}{n} }
\end{equation} As usual,  $g$ is the determinant of the matrix of components $g_{\mu\nu}$ of the metric tensor in a coordinate basis; $g$ is a scalar density of weight 2 and takes negative values because of the signature of the metric tensor.  $\hat{g}_{\mu\nu}$ is the conformal metric density.  The new variables  $\hat{g}_{\mu\nu}$  (or its inverse) and $\sqrt{-g}$ (or any nonzero power thereof) are those of Anderson and Finkelstein or are relevantly similar.   They further observe that  this scalar density is an absolute object in unimodular GTR. This observation seems unremarkable because that scalar density is not variational. For comparison, one recalls that Asher Peres rewrote the Lagrangian density for GTR in terms of the conformal metric density and a scalar density \cite{PeresPolynomial}; recently this idea was reinvented by M. O. Katanaev \cite{Katanaev}.  Surely the result is still GTR and not some other theory.   To my knowledge, no one (prior to Geroch, in effect) has ever considered whether the scalar density, even if varied in an action principle for GTR, might still count as an absolute object.  Once the question is raised about GTR with the Peres-type variables, a positive answer seems obvious: GTR has an absolute object!  This absolute object is a scalar density of nonzero weight, because every neighborhood in every model space-time admits coordinates (at least locally) in which the component of the scalar density has a value of $-1.$

   Interesting conclusions follow.  First, either Anderson's claim that GTR's novelty lay in its lack of absolute objects, or his analysis of absolute objects, is flawed.  Second, the scalar density is absolute despite being variational, somewhat as clock fields might be. Perhaps some people assume that any field varied in an action principle is dynamical (that is, not absolute),  even while  officially employing Anderson's definition of absoluteness.  Third, it would be useful to combine  hints from Anderson and Finkelstein about the (ir)reducibility of geometric objects with the  notion of equivalent geometric objects \cite{Nijenhuis} 
  to accommodate changes of basic variables or, as a field theorist might say, field redefinitions.  Finally, though some philosophers of physics profess to know absolute objects when they see them, even without an analysis, the case of GTR formulated using a conformal metric density and a scalar density  suggests otherwise.  Evidently no one has  spotted the absolute scalar density in GTR simply by inspection.  It follows that either one sometimes does not know an absolute object when one sees it, or that the Andersonian analysis of absolute objects gives the wrong answer for this example.  If the latter horn is accepted, then Peres's version of GTR in terms of a conformal metric density and a scalar density (both varied in the action principle) has no absolute object, whereas unimodular GTR in terms of a conformal metric density and a \emph{nonvariational} scalar density \emph{has} an absolute object, although the theories have the same geometric objects and nearly the same field equations (supplemented with N\"{o}ther identities). Such a claim requires justification.   Perhaps those who claim to spot absolute objects by inspection merely detect nonvariational objects in this instance?  Whether a theory has nonvariational  objects is, at least in some important examples, merely a question of its formulation, because tricks such as Rosen's Lagrange multiplier or the parametrization of preferred coordinates into clock fields can be employed to turn nonvariational fields into variational ones.\footnote{One hesitates to generalize too broadly  on this matter. In GTR in terms of the conformal metric density (or its inverse) and a scalar density, the latter counts as absolute, so one might be tempted not to vary it in the action principle.  But then the field equations are changed:  a cosmological constant enters as a constant of integration, as is well known.  The reason pertains to the mathematical form of the Lie derivative of a scalar density \cite{Israel}: for weight $w,$ $\pounds_{\xi} \phi=\xi^{\mu} \phi,_{\mu} + w \phi \xi^{\mu},_{\mu}$ and the $w$ term opens the door to the constant of integration.    For scalars, it makes no difference whether one varies   them as clock fields or not, because  the form of the generalized Bianchi identities \cite{Sundermeyer} and the linear independence of the gradients of the clock fields ensures that the same equations hold either way.  For Rosen's flat metric tensor trick, a new Lagrange multiplier field is introduced.   Thus the consequences of changing a nonvariational field into a variational one or \emph{vice versa} depend on which sort of geometric object is involved. This matter could use further study, perhaps with an eye on work on first-order and second-order actions for theories in which all fields are variational \cite{Ray,Lindstrom}. 
The absoluteness of the scalar density in GTR implies that it can be varied only at the cost of ceasing to  call GTR ``Lagrangian-based'' (\emph{c.f.} \cite{TLL}). }  If the having or lacking of absolute objects is merely a formal feature of a theory, then some new way of escaping the Kretschmann objection to the physical vacuity of general covariance \cite{NortonGC} is needed. Absent much healthy competition, the Andersonian project  is  worthy of attention even if its widely advertised diagnosis of the novelty of GTR is incorrect.

If the novelty of GTR does not consist in its lacking absolute objects (given Anderson's definition of them), still Anderson's project of analyzing the novelty of GTR might be fixable.  There are indeed interesting novel features of GTR that Anderson's framework uncovers or suggests.  For example, GTR apparently is novel in having an external symmetry group involving arbitrary functions of space and time and in having a group as large as the volume-preserving diffeomorphisms.  While Hiskes's proposal to invoke variational principles was too crude, some more sophisticated effort might succeed.  It is not presently clear whether it is best to admit that GTR has an absolute object or to redefine absolute objects to keep GTR from having any, if possible, but it seems worthwhile to consider the question.

\section{Conclusion}

Reviewing the Anderson-Friedman  absolute objects program and various possible counterexamples yields several conclusions.  First, eliminating irrelevant fields or portions thereof vindicates Friedman's resolution of the Jones-Geroch dust counterexample and apparently resolves the new tetrad-spinor counterexample.  Second, limitation of the mathematics to tensor fields has been detrimental by obscuring from view the tetrad-spinor and scalar density  cases, while leading to an overestimate of the force of Torretti's constant curvature spaces example. The mathematical theory of geometric objects is important for consideration of absolute objects. In particular, the geometric objects used should be irreducible.  Third, bringing into the philosophical discussion some neglected  physics literature sheds light on various issues.   Finally, the scalar density counterexample, which arguably is the only real problem for the Anderson-Friedman framework of the four considered here, shows that either GTR has an absolute object or the Andersonian  definition of absolute objects is flawed.

\section{Acknowledgments} The author thanks Don Howard, Jeremy Butterfield, Brandon Fogel, Roger Jones and David Malament for helpful  discussion and comments on the manuscript;  Harvey Brown and John Norton  for useful discussions;  A.  Camp, Ray Jensen and P. Nelson for bibliographic assistance; and, as noted above, Robert Geroch for suggesting the scalar density counterexample.



\end{document}